\documentstyle[mprocl]{article}

\input{psfig}

\def\lessapprox{\setbox0=\hbox{$<$}\setbox1=\hbox{$\sim$}
     \lower0.5\ht0
     \hbox{ \vbox{\baselineskip=0pt\lineskip=0.5pt\box0\box1} }}
\def\greatapprox{\setbox0=\hbox{$>$}\setbox1=\hbox{$\sim$}
     \lower0.5\ht0
     \hbox{ \vbox{\baselineskip=0pt\lineskip=0.5pt\box0\box1} }}

\begin{document}

\title{THE COSMIC FOAM:\\
\smallskip
STOCHASTIC GEOMETRY AND SPATIAL CLUSTERING\\
ACROSS THE UNIVERSE}

\author{Rien van de Weygaert}
\address{Kapteyn Institute, University of Groningen, Groningen, the Netherlands}


\maketitle\abstracts{
Galaxy redshift surveys have uncovered the existence of 
a salient and pervasive foamlike pattern in the distribution of 
galaxies on scales of a few up to more than a hundred Megaparsec. The 
significance of this frothy morphology of cosmic structure has been underlined 
by the results of computer simulations. These suggest the observed 
cellular patterns to be a prominent and natural aspect of cosmic structure 
formation for a large variety of scenarios within the context of the 
gravitational instability theory of cosmic structure formation.\\
\hskip 0.5truecm 
We stress the importance of stochastic geometry as a branch of mathematical 
statistics particularly suited to model and investigate nontrivial 
spatial patterns. One of its key concepts, Voronoi tessellations, represents 
a versatile and flexible mathematical model for foamlike patterns. Based on a 
seemingly simple definition, Voronoi tessellations define a wealthy stochastic 
network of interconnected anisotropic components, each of which can be identified 
with the various structural elements of the cosmic galaxy distribution. The  
usefulness of Voronoi tessellations is underlined by the fact that they appear 
to represent a natural asymptotic situation for a range of gravitational instability 
scenarios of structure formation in which void-like regions are prominent.\\
\hskip 0.5truecm 
Here we describe results of an ongoing thorough investigation of 
a variety of aspects of cosmologically relevant spatial distributions and statistics 
within the framework of Voronoi tessellations. Particularly enticing is the recent finding 
of a profound scaling of both clustering strength and clustering extent for the 
distribution of tessellation nodes, suggestive for the clustering properties 
of galaxy clusters. This is strongly suggestive of a hitherto 
unexpected fundamental and profound property of foamlike geometries. In a 
sense, cellular networks may be the source of an intrinsic  
``geometrically biased'' clustering.\\}

\section{Introduction}
Macroscopic patterns in nature are often due the
collective action of basic, often even simple, physical processes. 
These may yield a surprising array of complex and genuinely unique physical 
manifestations. The macroscopic organization into complex spatial patterns is 
one of the most striking. The rich morphology of such systems and patterns 
represents a major source of information on the underlying physics. This has 
made them the subject of a major and promising area of inquiry. However, most such 
studies still reside in a relatively youthful state of development, hampered by 
the fact that appropriate mathematical machinery for investigating and solidly 
characterizing the geometrical intricacies of the observed morphologies is not 
yet firmly in place.\\
\indent In an astronomical context one of the most salient geometrically complex 
patterns is that of the foamlike distribution of galaxies, revealed by 
a variety of systematic and extensive galaxy redshift surveys. Over the 
two past decades, these galaxy mapping efforts have gradually established the 
frothy morphology as a universal aspect of the spatial organization of matter 
in the Univers. Comprising features on a typical scale of tens of Megaparsec, it  
offers a direct link to the matter distribution in the primordial Universe. 
The cosmic web is therefore bound to contain a wealth of information on the cosmic 
structure formation process. It will therefore represent a key to unravelling one of 
the most pressing enigmas in modern astrophysics, the rise of the wealth and variety of 
structure in the present-day Universe from a almost perfectly smooth, virtually 
featureless, pristine cosmos. 

However, a lack of straightforward quantitative 
measures of such patterns has yet prevented a proper interpretation, or indeed 
identification, of all relevant pieces of information. Quantitative analysis of matter
distribution has been largely restricted to first order galaxy clustering
measures, useful in evaluating gross statistical properties of the matter
distribution but inept for characterizing  the intricate foamlike morphologies
observed on Megaparsec scales. 

Here we will address the meaning and interpretation of the cellular morphology 
of the cosmic matter distribution. Prominent as it is, its assessment rarely exceeds 
mere qualitative terminology, seriously impeding the potential exploitation of 
its content of significant information. One of the most serious omissions concerns 
a proper appreciation and understanding of the physical and statistical repercussions of 
the nontrivial cellular geometry. This propelled us to focus on this important aspect, for 
which we were impelled to invoke ideas and concepts from the relevant field of 
mathematics, stochastic geometry. Particularly fruitful has been our application and 
investigation of Voronoi tessellations, a central concept in this mathematical branch 
addressing the systematics of geometrical entities in a stochastic setting. The 
phenomological similarity of Voronoi foams to the cellular morphology seen in the 
galaxy distribution justifies further exploration of its virtues as a model 
for cosmic structure. In the following we will indicate that such similarity 
is a consequence of the tendency of gravity to shape and evolve structure emerging 
from a random distribution of tiny density deviations into a network of anisotropically 
contracting features. Its application gets solidly underpinned by a thorough 
assessment of the implications for spatial clustering, vindicating the close 
resemblance of Voronoi foams to the frothy patterns in the observed reality. 
It is within the context of testing its spatial statistical properties that 
unexpected profound `scaling' symmetries were uncovered, shedding new light 
on the issue of ``biased'' spatial clustering.\\

\section{Patterns in the Galaxy Distribution: the Cosmic Foam.}
One of the most striking examples of a physical system displaying a salient
geometrical morphology, and the largest in terms of sheer size, is 
the Universe as a whole. The past few decades have revealed  that on scales of a 
few up to more than a hundred Megaparsec, galaxies conglomerate into intriguing 
cellular or foamlike patterns that pervade throughout the observable cosmos. 
A dramatic illustration is the map of the 2dF Galaxy Redshift Survey and 
the newest results of the SDSS survey (see contribution M. Strauss). The 
recently published map of the distribution of more than 150,000 galaxies 
in a narrow region on the sky yielded by the 2dF -- two-degree field -- redshift survey.
Instead of a homogenous distribution, we recognize a sponge-like arrangement, with 
galaxies aggregating in filaments, walls and nodes on the periphery of giant voids. 

\begin{figure}
\vskip -1.0truecm
\centering\mbox{\hskip -0.2truecm\psfig{figure=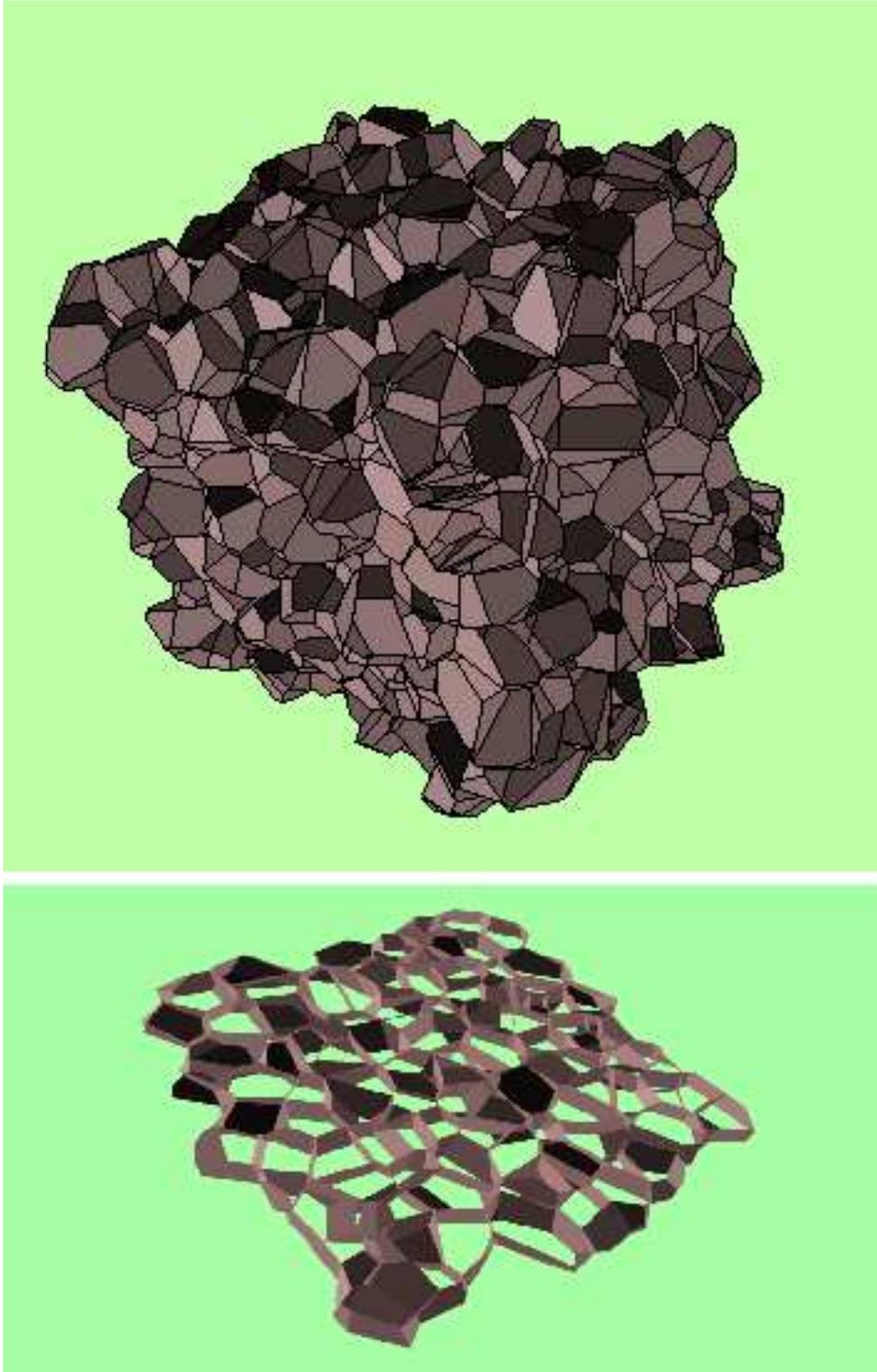,width=13.0cm}}
\caption{A full 3-D tessellation comprising 1000 Voronoi cells/polyhedra 
generated by 1000 Poissonian distributed nuclei. Courtesy: Jacco Dankers}
\end{figure}
This frothy geometry of the Megaparsec Universe is evidently one of the most 
prominent aspects of the cosmic fabric, outlined by galaxies populating huge 
{\it filamentary} and {\it wall-like} structures, the sizes of the most conspicuous 
one frequently exceeding 100h$^{-1}$ Mpc. The closest and best studied of these 
massive anisotropic matter concentrations can be identified with known supercluster 
complexes, enormous structures comprising one or more rich clusters of galaxies and 
a plethora of more modestly sized clumps of galaxies. A prominent and representative 
nearby specimen is the Perseus-Pisces supercluster,  a 5$h^{-1}$ wide ridge of 
at least 50$h^{-1}$ Mpc length, possibly extending out to a total length of 
140$h^{-1}$ Mpc. In addition to the presence of such huge filaments the galaxy distribution 
also contains vast planar assemblies. A striking example of is the {\it Great Wall}, 
a huge planar assembly of galaxies with dimensions that are estimated to be of the 
order of $60h^{-1} \times 170h^{-1} \times 5h^{-1}$ Mpc (Geller \& Huchra 1989). Within and 
around these anisotropic features we find a variety of density condensations, ranging from 
modest groups of a few galaxies up to massive compact {\it galaxy clusters}. The latter 
stand out as the most massive fully collapsed and virialized objects in the 
Universe. In nearby representatives like the Virgo and Coma cluster typically 
more than a thousand galaxies have been identified within a radius of a mere 
1.5$h^{-1}$ Mpc around the core. They may be regarded as a particular population of cosmic 
structure beacons as they typically concentrate near the interstices of the 
cosmic web, {\it nodes} forming a recognizable tracer of the cosmic matter distribution out 
to vast distances (e.g. Borgani \& Guzzo 2001). Complementing this cosmic inventory 
leads to the existence of large {\it voids}, enormous 
regions with sizes in the range of $20-50h^{-1}$ Mpc that are practically devoid of 
any galaxy, usually roundish in shape. The earliest recognized one, the 
Bo\"otes void (Kirshner et al. 1981, 1987), a conspicuous almost 
completely empty spherical region with a diameter of 
around $60h^{-1}$Mpc, is still regarded as the canonic example. The role of voids 
as key ingredients of the cosmic matter distribution has since been 
convincingly vindicated in various extensive redshift surveys, up to the recent results 
produced by 2dF redshift survey and the Sloan redshift surveys.

Of utmost significance for our inquiry into the issue of cosmic structure formation is 
the fact that the prominent structural components of the galaxy distribution -- clusters, 
filaments, walls and voids -- are not merely randomly and independently 
scattered features. On the contrary, they have arranged themselves in a seemingly 
highly organized and structured fashion, the {\it cosmic foam}. They are woven into 
an intriguing {\it foamlike} tapestry that permeates the whole of the 
explored Universe. Voids are generically associated with surrounding density 
enhancements. In the galaxy distribution they represent both contrasting as well as 
complementary components ingredients, the vast under-populated regions, (the {\it voids}), 
being surrounded by {\it walls} and {\it filaments}. At the intersections of the latter 
we often find the most prominent density enhancements in our universe, the 
{\it clusters} of galaxies. 
\section{Gravitational Foam Formation and Bubble Dynamics.}
Foamlike patterns have not only been confined to the real world. Equally important 
has been the finding that foamlike patterns do occur quite naturally in a vast 
range of structure formation scenarios within the context of the generic framework of 
gravitational instability theory. Prodded by the steep increase in computing 
power and the corresponding proliferation of ever more sophisticated and extensive 
simulation software, a large range of computer models of the structure formation process 
have produced telling images of similar foamlike morphologies. They reveal an 
evolution proceeding through stages characterized by matter 
accumulation in structures with a pronounced cellular morphology. 

The generally accepted theoretical framework for the formation of structure is that of 
gravitational instability. The formation and moulding of structure is ascribed to the 
gravitational growth of tiny initial density- and velocity deviations from the global 
cosmic density and expansion. An important aspect of the gravitational formation 
process is its inclination to progress via stages in which the cosmic matter 
distribution settles in striking anisotropic patterns. Aspherical overdensities, on 
any scale and in any scenario, will contract such that they become increasingly 
anisotropic, as long as virialization has not yet set in. At first they turn into 
a flattened `pancake', possibly followed by contraction into an elongated filament. Such 
evolutionary stages precede the final stage in which a virialized object,  
e.g. a galaxy or cluster, will emerge. This tendency to collapse anisotropically 
finds its origin in the intrinsic primordial flattening of the overdensity, augmented 
by the anisotropy of the gravitational force field induced by the external matter 
distribution, i.e. by tidal forces. Naturally, the induced anisotropic collapse 
has been the major agent in shaping the cosmic foamlike geometry. 

Inspired by early computer calculations, Icke (1984) pointed out that for the 
understanding of the formation of the large coherent patterns pervading the 
Universe it may be more worthwhile to direct attention to the complementary  
evolution of underdense regions. By contrast to the overdense features, the 
low-density regions start to take up a larger and larger part of the volume of 
the Universe. Icke (1984) then made the interesting observation 
that the arguments for the dynamics and evolution of slightly anisotropic -- 
e.g. ellipsoidal -- primordial overdensities are equally valid when considering 
the evolution of {\it low\/}-density regions. The most important difference is 
that the sense of the final effect is reversed. The continuously stronger anisotropy  
of the force field in collapsing ellipsoidal leads to the characteristics tendency for 
{\it slight initial asphericities to get amplified during the collapse}, the major 
internal mechanism for the formation of the observed filaments in the galaxy 
distribution. By contrast, a void is effectively a region of negative density in a 
uniform background. Therefore, they will expand as the overdense regions collapse, 
while {\it slight asphericities decrease as the voids become larger}. This can be 
readily appreciated from the fact that with respect to an equally deep spherical 
underdensity, an ellipsoidal void has a decreased rate of expansion along 
the longest axis of the ellipsoid and an increased rate of expansion along 
the shortest axis. Together with the implied {\it Hubble-type velocity field}, 
voids will thus behave like low-density `super-Hubble' expanding patches in the 
Universe. To describe this behaviour we coined the term  
``Bubble Theorem'' (Icke 1984). 

Evidently, we have to be aware of the serious limitations of the ellipsoidal model. 
It grossly oversimplifies in disregarding important aspects like the 
presence of substructure in and the immediate vicinity of peaks and dips 
in the primordial density field. Still, it is interesting to realize that 
in many respects the homogeneous model is a better approximation 
for underdense regions than it is for overdense ones. Voids expand and 
get drained, and the interior of a (proto)void rapidly flattens out, 
which renders the validity of the approximation accordingly better.
Such behaviour was clearly demonstrated in circumstances of voids embedded in 
a full complex general cosmic density field (see e.g. Van de Weygaert \& van 
Kampen 1993, their Fig. 16). Their systematic study also showed how voids in 
general will evolve towards a state in which they become genuine 
{\it ``Superhubble Bubbles''}. 

In realistic circumstances, expanding voids will sooner or later encounter 
their peers or run into dense surroundings. The volume of space available to 
a void for expansion is therefore restricted. Voids will also be influenced by the 
external cosmic mass distribution, and substructure may represent an additional 
non-negligible factor within the void's history. In general, we deal with a 
complex situation of a field of expanding voids and collapsing peaks, of voids and 
peaks over a whole range of sizes and masses, expanding at different rates and 
at various stages of dynamical development. For the purpose of our geometric 
viewpoint, the crucial question is whether it is possible to identify some 
characteristic and simplifying elements within such a complex. Indeed, simulations 
of void evolution (e.g. Dubinski et al. 1993) represent a suggestive illustration of 
a hierarchical process akin to the {\it void hierarchy} seen in realistic 
simulations (e.g. Van de Weygaert 1991b). It shows the maturing of small-scale voids 
until their boundaries would reach a shell-crossing catastrophe, after which they merge 
and dissolve into a larger embedding void. This process gets continuously repeated 
as the larger parent voids in turn dissolve into yet larger voids. For a primordial 
Gaussian density field, corresponding analytical calculations 
(Sheth \& Van de Weygaert 2002) then yield a void size distribution (broadly) 
peaked around a characteristic void size.  

A bold leap then brings us to a geometrically interesting situation. Taking the 
voids as the dominant dynamical component of the Universe, and following 
the ``Bubble Theorem'', we may think of the large scale structure as a close 
packing of spherically expanding regions. Then, approximating a peaked 
void distribution by one of a single scale, we end up with a situation 
in which the matter distribution in the large scale Universe is set up 
by matter being swept up in the bisecting interstices between spheres 
of equal expansion rate. This {\it ASYMPTOTIC} description of the 
cosmic clustering process leads to a geometrical configuration that 
is one of the main concepts in the field of stochastic geometry: {\it 
VORONOI TESSELLATIONS.}
\section{Voronoi Tessellations: the Geometric Concept}
A Voronoi tessellation of a set of nuclei is a space-filling
network of polyhedral cells, each of which delimits that part of space
that is closer to its nucleus than to any of the other nuclei.
In three dimensions a Voronoi foam consists of a packing of Voronoi cells,
each cell being a convex polyhedron enclosed by the bisecting planes between
the nuclei and their neighbours. A Voronoi foam consists of four 
geometrically distinct elements: the polyhedral cells ({\it 
voids\/}), their walls ({\it pancakes\/}), edges ({\it filaments\/}) 
where three walls intersect, and nodes ({\it clusters\/}) where four 
filaments come together.
\begin{figure}[t]
\vskip -3.0truecm
\centering\mbox{\psfig{figure=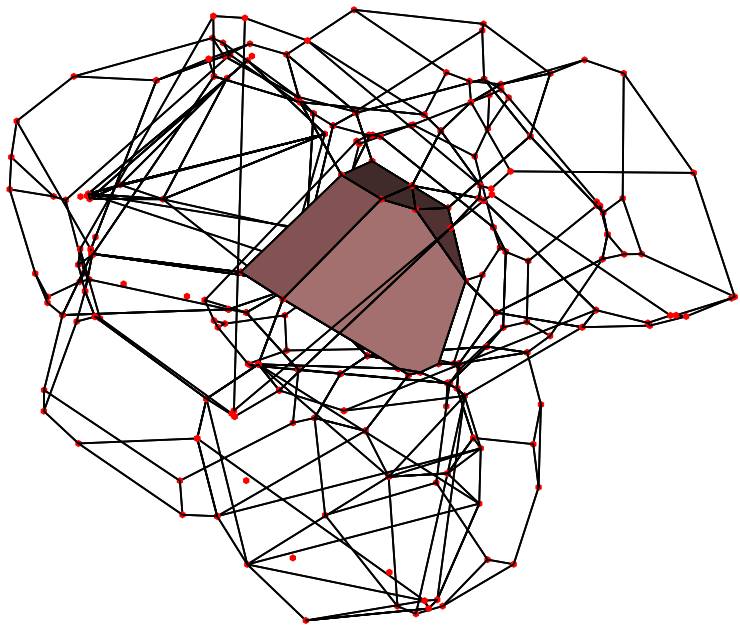,angle=90.0,width=10.0cm}}
\vskip -1.0truecm
\caption{Wireframe illustration of interrelation between various Voronoi 
tessellation elements. The central ``Voronoi cell'' is surrounded by its wire-frame 
depicted ``contiguous'' Voronoi neighbours. The boundaries of the cells 
are the polygonal ``Voronoi walls''. The wire edges represent the 
Voronoi edges. The ``Voronoi vertices'', indicated by red dots, are 
located at each of the 2 tips of a Voronoi edge, each of them located at 
the centre of the circumsphere of a corresponding set of four nuclei. 
Courtesy: Jacco Dankers.}
\vskip -0.5truecm
\end{figure}
Formally, each Voronoi region $\Pi_i$ is the set of points which is nearer to nucleus $i$ 
than to any of the other nuclei $j$ in a set $\Phi$ of nuclei $\{x_i\}$ in
$d$-dimensional space $\Re^d$, or a finite region thereof,
$\Pi_i = \{{\vec x} \vert d({\vec x},{\vec x}_i) < d({\vec x},{\vec x}_j)\ ,
\ \forall\ j \not= i \}$, where ${\vec x}_j$ are the position vectors of the nuclei 
in $\Phi$, and $d({\vec x},{\vec y})$ 
the Euclidian distance between ${\vec x}$ and ${\vec y}$ (evidently, one can extend the 
concept to any arbitrary distance measure). From this basic definition, we can 
directly infer that each Voronoi region $\Pi_i$ is the intersection of the open 
half-spaces bounded by the perpendicular bisectors (bisecting planes in 3-D) of 
the line segments joining the nucleus $i$ and any of the the other nuclei. This  
implies a Voronoi region $\Pi_i$ to be a convex polyhedron (or polygon when in 2-D), 
a {\it Voronoi polyhedron}. The complete set of $\{\Pi_i\}$ constitute a space-filling 
tessellation ofmutually disjunct cells in d-dimensional space $\Re^d$, the {\it Voronoi
tessellation} ${\cal V}(\Phi)$ relative to $\Phi$. A good impression of the morphology of 
a complete Voronoi tessellation can be seen in figure 1, a tessellation of 1000 cells
generated by a Poisson distribution of 1000 nuclei in a cubic box.

Taking the three-dimensional tessellation as the archetypical representation
of structures in the physical world, we can identify four constituent {\it elements} 
in the tessellation, intimately related aspects of the full Voronoi tessellation. 
In addition to (1) the polyhedral {\it Voronoi cells} $\Pi_i$ these are (2) the polygonal
{\it Voronoi walls} $\Sigma_{ij}$ outlining the surface of the Voronoi cells,
(3) the one-dimensional {\it Voronoi edges} $\Lambda_{ijk}$ defining the rim of
both the Voronoi walls and the Voronoi cells, and finally (4) the
{\it Voronoi vertices} $V_{ijkl}$ which mark the limits of edges, walls and cells. 
While each Voronoi cell is defined by one individual nucleus in the
complete set of nuclei $\Phi$, each of the polygonal Voronoi walls 
$\Sigma_{ij}$ is defined by two nuclei $i$ and $j$, consisting of points ${\vec x}$ having
equal distance to $i$ and $j$. The Voronoi wall $\Sigma_{ij}$ is the 
subregion of the full bisecting plane of $i$ and $j$ which consists 
of all points ${\vec x}$ closer to both $i$ and $j$ than
other nuclei in $\Phi$. In analogy to the definition of a
Voronoi wall, a Voronoi edge $\Lambda_{ijk}$ is a subregion of the 
equidistant line defined by three nuclei $i$, $j$ and $k$, the subregion consisting of
all points ${\vec x}$ closer to $i$, $j$ and $k$ than to any of the other
nuclei in $\Phi$. Evidently, it is part of
the perimeter of three walls as well, $\Sigma_{ij}$, $\Sigma_{ik}$ and
$\Sigma_{jk}$.  Pursuing this enumeration, Voronoi vertices $V_{ijkl}$
are defined by four nuclei, $i$, $j$, $k$ and $l$, being the one point
equidistant to these four nuclei and closer to them than to any of the other nuclei
belonging to $\Pi_i$. Note that this implies that the circumscribing
sphere defined by the four nuclei does not contain any other nuclei. 
To appreciate the interrelation between these different geometric aspects, 
figure 2 lifts out one particular Voronoi cell from a clump of a dozen Voronoi cells. 
The central cell is the one with its polygonal Voronoi walls surface-shaded,
while the wire-frame representation of the surrounding Voronoi cells reveals the Voronoi 
edges defining their outline and the corresponding vertices as red dots. Notice, 
how the distribution of vertices, generated by the stochastic point process 
of nuclei, is in turn a new and uniquely defined point process, that of the {\it 
vertices} !!!

\section{Voronoi Tessellations: the Cosmological Context}
In the cosmological context {\it Voronoi Tessellations} represent the {\it 
Asymptotic Frame} for the ultimate matter distribution distribution in any 
cosmic structure formation scenario, the skeleton delineating the destination of the 
matter migration streams involved in the gradual buildup of cosmic structures. 
The premise is that some primordial cosmic process generated a density fluctuation 
field. In this random density field we 
can identify a collection of regions where the density is slightly less 
than average or, rather, the peaks in the primordial gravitational potential perturbation 
field. As we have seen, these regions are the seeds of the voids. These underdense patches 
become ``expansion centres'' from which matter flows away until it runs into its 
surroundings and encounters similar material flowing out of adjacent voids. 
Notice that the dependence on the specific structure formation scenario at hand is 
entering via the spatial distribution of the sites of the density dips in the 
primordial density field, whose statistical properites are fully determined
by the spectrum of primordial density fluctuations.

Matter will collect at the interstices between the expanding voids. In the 
asymptotic limit of the corresponding excess Hubble parameter being the 
same in all voids, these interstices are the bisecting planes, perpendiculary 
bisecting the axes connecting the expansion centres. For any given set of expansion 
centres, or {\it nuclei}, the arrangement of these planes define a unique process 
for the partitioning of space, a {\it Voronoi tessellation\/} (Voronoi 1908, see 
Fig. 1 and 2). A particular realisation of this process (i.e. a specific 
subdivision of $N$-space according to the Voronoi tessellation) may be called 
a {\it Voronoi foam\/} (Icke \& Van de Weygaert 1987). 
Within such a cellular framework the interior of each ``{\it VORONOI CELL}'' is 
considered to be 
a void region. The planes forming the surfaces of the cells 
are identified with the ``{\it WALLS}'' in the galaxy distribution (see e.g. Geller 
\& Huchra 1989). The ``{\it EDGES}'' delineating the rim of each wall are to be  
identified with the filaments in the galaxy distribution. In general, what is 
usually denoted as a flattened ``supercluster'' or cosmic ``wall'' will 
comprise an assembly of various connecting walls in the Voronoi foam. 
The elongated ``superclusters'' 
or ``filaments'' will usually consist of a few coupled edges (Fig. 3 clearly 
illustrates this for the Voronoi kinematic model). Finally, the 
most outstanding structural elements are the ``{\it VERTICES}'', tracing the surface of 
each wall, outlining the polygonal structure of each wall and limiting the 
ends of each edge. They correspond to the very dense compact nodes within the cosmic 
network, amongst which the rich virialised Abell clusters form the most 
massive representatives. 

Cosmologically, the great virtue of the Voronoi foam is that it provides a
conceptually simple model for a cellular or foamlike distribution of
galaxies, whose ease and versatility of construction makes it an ideal
tool for statistical studies. By using such geometrically constructed models 
one is not restricted by the resolution or number of particles. A 
cellular structure can be generated over a part of space beyond the 
reach of any N-body experiment. Even though the model does not and cannot addres
the galaxy distribution on small scales, it is nevertheless a useful 
prescription for the spatial distribution of the walls and filaments 
themselves. This makes the Voronoi model particularly suited
for studying the properties of galaxy clustering in cellular structures on
very large scales, for example in very deep pencil beam surveys, and for
studying the clustering of clusters in these models. 
\section{Voronoi Galaxy Distributions}
Having established the cosmological context for Voronoi tessellations in the form 
of, approximate and asymptotic, skeletal template for the large-scale mass 
distribution we set about to generate the corresponding matter distributions. Matter 
is supposed to aggregate in and around the various geometrical elements of the cosmic 
frame, such as the walls, the filaments and the vertices.

It is the stochastic yet non-Poissonian geometrical distribution of the walls, 
filaments and clusters embedded in the cosmic framework which generates 
the large-scale clustering properties of matter and the related galaxy populations. 
The small-scale distribution of galaxies, i.e. the distribution within the various 
components of the cosmic skeleton, will involve the complicated details of highly 
nonlinear small-scale interactions of the gravitating matter. N-body simulations are 
preferred for treating that problem. For our purposes, we take the route of 
complementing the large-scale cellular distribution induced by Voronoi patterns by 
a user-specified small-scale distribution of galaxies. Ideally, well-defined and elaborate 
physical models would fill in this aspect. A more practical alternative approch involves the 
generation of either tailor-made purely heuristic ``galaxy'' distributions in and 
around the various elements of a Voronoi tessellation (e.g. pure uniform distributions).  
Alternatively, we can generate distributions that more closely resemble the outcome of dynamical 
simulations, and represent an idealized and asymptotic description thereof. Such a model is the 
{\it kinematic model} defined by Van de Weygaert \& Icke (1989). 
\begin{figure}[t]
\vskip -0.5truecm
\centering\mbox{\hskip -0.2truecm\psfig{figure=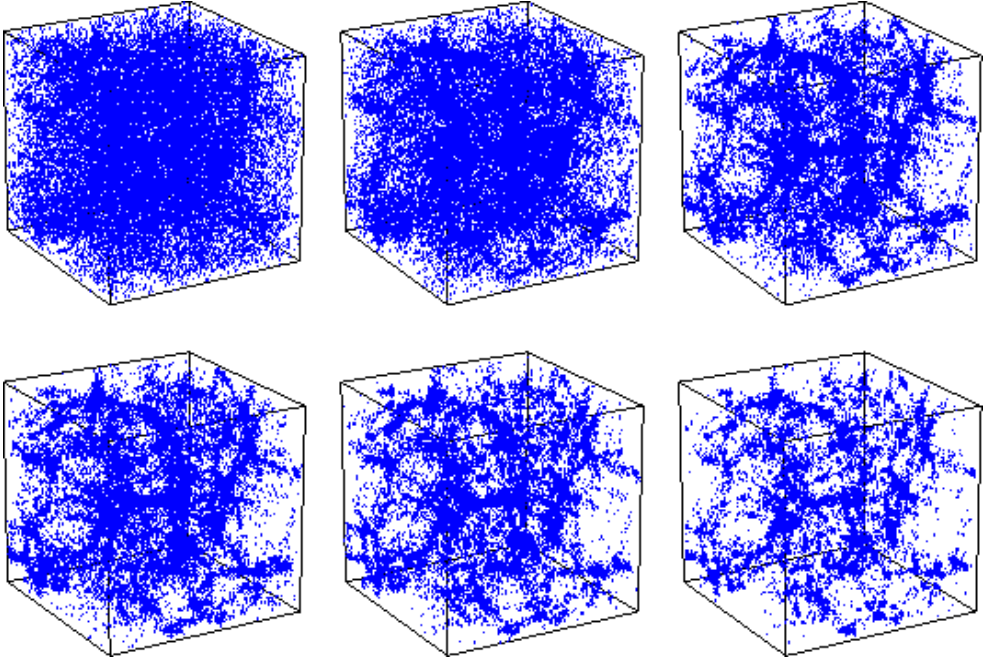,width=13.0cm}}
\caption{A sequel of consecutive timesteps 
within the kinematic Voronoi cell formation process. The depicted boxes  
have a size of $100h^{-1} \hbox{Mpc}$. Within these cubic volumes some $64$ Voronoi 
cells with a typical size of $25h^{-1}\hbox{Mpc}$ delineate the cosmic framework around 
which some 32000 galaxies have aggregated (corresponding roughly  
to the number density of galaxies yielded by a Schechter luminosity 
function with parameters according to Efstathiou, Ellis \& Peterson 1988), where 
we restricted ourselves to galaxies brighter than $M_{gal} = -17.0$. In the 
full ``simulation box'' of $200h^{-1}\hbox{Mpc}$, this amounts to 268,235 galaxies.}
\vskip -0.5truecm
\end{figure}
Particular emphasis should be put on that fact that this Voronoi strategy has the unique and fundamental feature of studying galaxy distributions around geometrical features 
that themselves have some distinct and well-defined stochastic spatial distribution. 
The galaxies are residing in walls, filaments and vertices which are distributed 
themselves as an integral component of the Voronoi spatial network. Their distribution 
is not a pure random, but instead one in which these components themselves are 
spatially strongly correlated, connecting into coherent ``super''structures !!! 
This background frame of spatially clustered geometrical elements not only 
determines the overall clustering properties of its galaxy population, it also 
represents and distinguishes it from from less physically motivated stochastic 
toy models (e.g. the double Poisson process). 

\subsection{Voronoi galaxy distributions: the Kinematic Model}
The kinematic Voronoi model is based on the notion that when matter streams out 
of the voids towards the Voronoi skeleton, cell walls form when material from one void 
encounters that from an adjacent one. In the original ``pancake picture'' of Zel'dovich and 
collaborators, it was gaseous dissipation fixating the pancakes (walls), 
automatically leading to a cellular galaxy distribution. But also when the 
matter is collisionless, the walls may be hold together by their own self-gravity. 
Accordingly, the structure formation scenario of the kinematic model proceeds as follows. 
Within a void, the mean distance between galaxies increases uniformly in the course 
of time. When a galaxy tries to enter an adjacent cell, the gravity of the wall, aided 
and abetted by 
dissipational processes, will slow its motion down. On the average, this amounts to 
the disappearance of its velocity component perpendicular to the cell wall. Thereafter, the
galaxy continues to move within the wall, until it tries to enter the
next cell; it then loses its velocity component towards that cell, so
that the galaxy continues along a filament. Finally, it comes to rest
in a node, as soon as it tries to enter a fourth neighbouring void. 
Of course the full physical picture is expected to differ considerably in 
the very dense, highly nonlinear regions of the network, around the filaments 
and clusters. Nonetheless, the Voronoi kinematic model produces 
a structural morphology containing the revelant characteristics of the 
cosmic foam, both the one seen in large redshift surveys as the one found in 
the many computer model N-body simulations. 

The evolutionary progression within our Voronoi kinematic scheme, from an almost 
featureless random distribution, via a wall-like and filamentary morphology towards 
a distribution in which matter ultimately aggregates into conspicuous compact 
cluster-like clumps can be readily appreciated from the sequence of 6 cubic 3-D 
particle distributions in Figure 3. The steadily increasing contrast of the various structural features is accompanied by a gradual shift in topological nature of the 
distribution. The virtually uniform particle 
distribution at the beginning (upper lefthand frame) ultimately unfolds into the 
highly clumped distribution in the lower righthand frame. At first only a faint 
imprint of density enhancements and depressions can be discerned. In the subsequent 
first stage of nonlinear evolution we see a development of the matter distribution 
towards a wall-dominated foam. The contrast of the walls with respect to the general 
field population is rather moderate (see e.g. second frame), and most obviously 
discernable by tracing the sites where the walls intersect and the galaxy density is 
slightly enhanced. The ensuing frames depict the gradual progression via a wall-like 
through a filamentary towards an ultimate cluster-dominated matter distribution. By 
then nearly all matter has streamed into the nodal sites of the cellular network. The 
initially almost hesitant rise of the clusters quickly turns into a strong and 
incessant growth towards their appearance as dense and compact features which 
ultimately stand out as the sole dominating element in the cosmic matter 
distribution (bottom righthand frame). 

\section{Superclustering: the clustering of clusters}
Maps of the spatial distribution of clusters of galaxies show that clusters 
themsvelves are not Poissonian distributed, but turn out to 
be highly clustered (see e.g. Bahcall 1988). They aggregate to form huge 
supercluster complexes. For the sake of clarity, it is worthwhile to notice that 
such superclusters represent moderate density enhancements on a scale 
of tens of Megaparsec, typically in the order of a few times the average. 
They are still co-expanding with the Hubble flow, be it at a slightly  
decelerated rate, and are certainly not to be compared with collapsed, let 
alone virialized, identifiable physical entities like clusters.
\subsection{Measuring Superclustering}
For the context of the present work, which is specifically intended to shed 
light on the relation between clusters and the cosmic network, we stress 
the common quantification of clustering, in conjunction with three particularly 
relevant aspects following from these analyses. By folklore, cosmologists tend 
to concentrate on quantifying the 
clustering of galaxies merely through the first order deviations from a 
featureless purely uniform Poisson istribution of points. Figuring prominently 
as th\'e ``tool of the trade'' in these statistical studies 
is the two-point correlation function $\xi(r)$ (and its sky-projected equivalent, 
the angular two-point correlation function $\omega(\theta)$), 
\begin{equation}
dP(r)\,=\,{\bar n}^2\,(1 + \xi(r))\,dV_1\,dV_2\,,
\end{equation}
\noindent quantifying the excess probability of finding a pair of points in volume 
elements $d V_1$ and $d V_2$ separated by a distance $r$ in a point 
sample of average number density ${\bar n}$. For the sake of terminology, we wish 
to point out that in the field of cosmology the amplitude of $\xi(r)$ is traditionally 
expressed in terms of the scale $r_{\rm o}$, $\xi(r_{\rm o})=1$, which then is usually 
indicated by the name ``correlation length''. However, 
in the following we will use the more correct name of ``clustering length''. 
Rather than a characteristic geometric scale, $r_{\rm o}$ is more a 
measure for the ``compactness'' of the spatial clustering, set mainly by the 
small-scale clustering. A more significant scale within the context of
the geometry of the spatial patterns in the density distribution is 
the scale at which $\xi(r_{\rm a})=0$. As a genuine scale of coherence, we reserve 
the name ``correlation length'' for this scale. For the morphology of the 
nontrivial spatial structures, the target of our study, it is a highly meaningful measure. 
\subsection{Distinctive Superclustering properties} 
The {\it first} characteristic of superclustering is the finding that the clustering 
of clusters is considerably more pronounced than that of galaxies. According to 
most studies the two-point correlation function $\xi_{cc}(r)$ of clusters is consistent 
with it being a scaled version of the power-law galaxy-galaxy correlation function, 
$\xi_{cc}(r)=(r_{\circ}/r)^\gamma$. While most agree on the same 
slope $\gamma \approx 1.8$ and a correlation amplitude that is 
significantly higher than that for the galaxy-galaxy correlation function, the 
estimates for the exact amplitude differ considerably from a factor 
$\simeq 10-25h^{-1}\hbox{Mpc}$. The original value found for the 
``clustering length'' $r_{\rm o}$ for 
rich $R\geq 1$ Abell clusters was $r_{\rm o} \approx 25h^{-1}\hbox{Mpc}$ 
(Bahcall \& Soneira 1983), up to a scale of $100 h^{-1}$ Mpc (Bahcall 1988). Later 
work favoured more moderate values in the order of $15-20h^{-1}\hbox{Mpc}$ 
(e.g. Sutherland 1988, Dalton et al. 1992, Peacock \& West 1992). 
\begin{figure}[t]
\vskip -1.0truecm
\centering\mbox{\hskip -1.truecm\psfig{figure=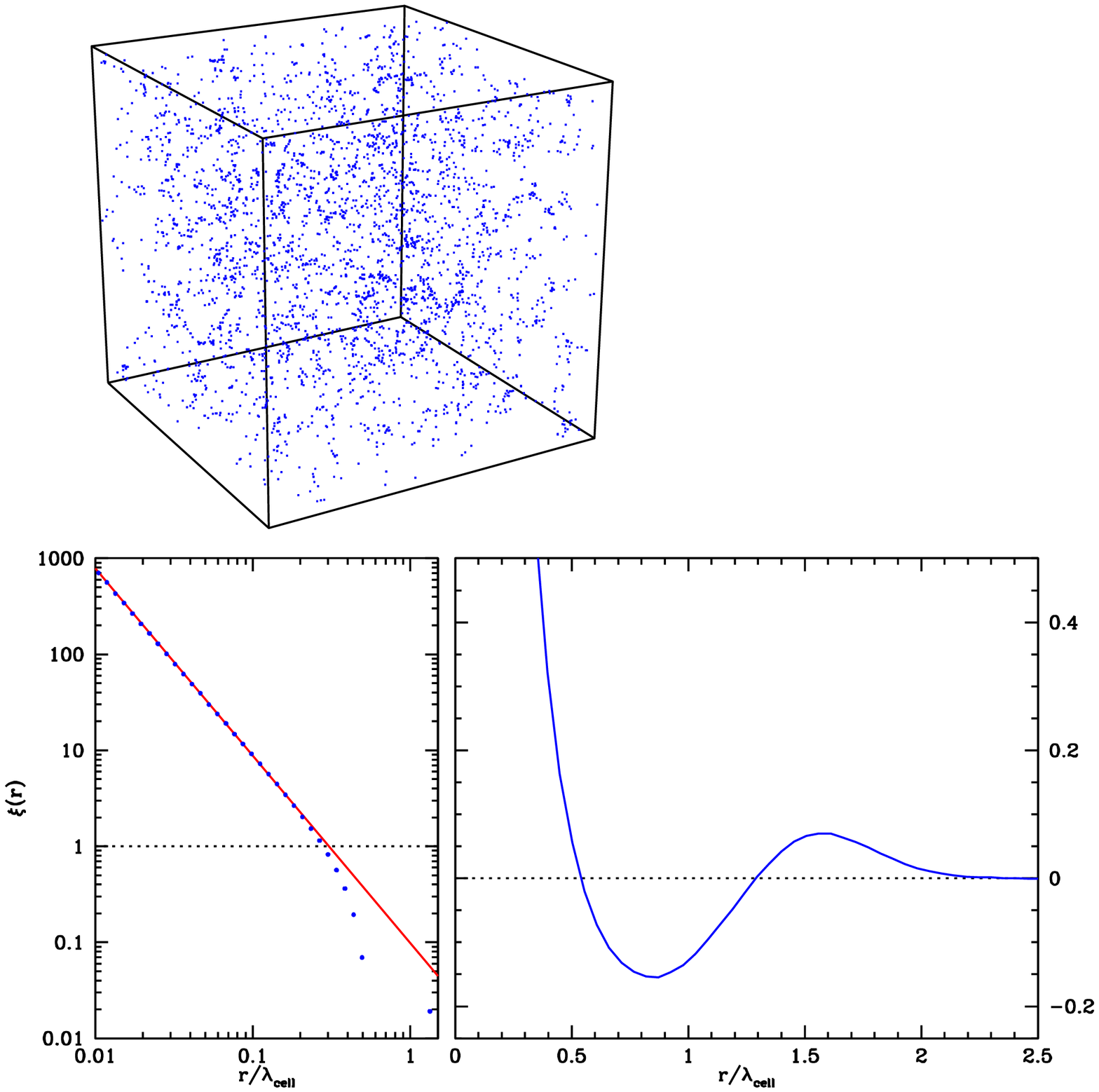,width=14.5cm}}
\vskip -0.3truecm
\end{figure}
\begin{figure}
\caption{Two-point correlation function analysis of a selection of galaxies in 
a Voronoi kinematic model realization. Top frame: a spatial 3-D depiction of a full 
galaxy sample in a box of size $150h^{-1}\hbox{\rm Mpc}$, at a stage corresponding 
to the present cosmic epoch $\sigma(8h^{-1}\hbox{\rm Mpc}\approx 1$. The cellular 
morphology with walls and filaments forms a marked pattern throughout the box, with 
sites of a few conspicuously dense cluster ``nodes'' standing out. 
Bottom left: a log-log plot of the $\xi(r)$, with distance $r$ in units 
of the basic cellsize $\lambda_{\rm cell}$. The power-law character of $\xi$ 
up to $r \sim 0.5 \lambda_c$ is evident. Bottom right: a lin-lin plot of 
$\xi$. The beautiful ringing behaviour out to scales $r \sim 2 \lambda_{\rm cell}$ 
has been amply recovered. From: Van de Weygaert 2002.}
\vskip -0.5truecm
\end{figure}

A related {\it second} characteristic of superclustering is that the differences in 
estimates of $r_{\rm o}$ are at least partly related to the specific selection of 
clusters, i.e. the applied definition of clusters. Studies dealt with 
cluster samples of rich $R\geq 1$ Abell clusters, others also included 
poorer clusters, or employed a physically well-founded 
criterion on the basis of X-ray emission. On the basis of such 
analyses we find a trend of an increasing clustering strength as the 
clusters in the sample become more rich ($\approx$ massive). On the 
basis of the first related studies, Szalay \& Schramm (1985) even 
put forward the (daring) suggestion that samples of clusters selected 
on richness would display a `fractal' clustering behaviour, in which the 
clustering scale $r_{\rm o}$ would scale linearly with the typical scale $L$ 
of the cluster catalogue. This typical scale $L(R)$ is then the mean 
separation between the clusters of richness higher than $R$: 
$\xi_{cc}(r)=\beta\ (L(r)/r)^\gamma$ where $L(R)=n^{-1/3}$.  
While the exact scaling of $L(r)$ with mean number density $n$ is questionable, 
observations follow the qualitative trend of a monotonously increasing $L(R)$. 
It also seems to adhere to the increasing level of clustering that selections of 
more massive clusters appear to display in large-scope N-body simulations 
(e.g. Colberg 1998), given some telling detailed differences.

A {\it third} aspect of superclustering, one that often escapes emphasis but 
which we feel is important to focus attention on, is the issue of the spatial range over 
which clusters show positive correlations, the ``coherence'' scale of cluster 
clustering. Currently 
there is ample evidence that $\xi_{cc}(r)$ extends out considerably further 
than the galaxy-galaxy correlation $\xi_{gg}$, possibly out to 
$50h^{-1}-100h^{-1}\hbox{Mpc}$. This is not in line with 
conventional presumption that the stronger level of cluster clustering 
is due to the more clustered locations of the (proto)cluster peaks in the 
primordial density field with respect to those of (proto)galaxy peaks. According 
to this conventional ``peak bias'' scheme we should not find significant 
non-zero cluster-cluster correlations on scales where the galaxies no 
longer show any significant clustering. If indeed $\xi_{gg}$ is negligible  
on these large scales, explaining the large scale cluster-cluster clustering 
may be posing more complications than a simple interpretation would suggest.
\section{Superclustering: the Voronoi Vertex Distribution}
In the Voronoi description vertices are identified with the clusters of galaxies, 
a straightforward geometric identification without need to invoke additional 
descriptions. Like genuine clusters, these vertices then act as the condensed and 
compact complexes located at the interstices in the cosmic framework. 
The immediate and highly significant consequence is that -- for a given Voronoi 
foam realization -- the spatial distribution of clusters is fully and uniquely 
determined. The study of the clustering of these vertices can therefore be done 
without any further assumptions, fully set by the geometry of the tessellation. 
When doing this, we basically use the fact that {\it the Voronoi node
distribution is a topological invariant\/} in co-moving coordinates,
and does not depend on the way in which the walls, filaments, and
nodes are populated with galaxies. The statistics of the
nodes should therefore provide a robust measure of the Voronoi properties.

A first inspection of the spatial distribution of Voronoi vertices (Fig. 4, 
top frame) immediately reveals that it is not a simple random Poisson 
distribution. The full spatial distribution of Voronoi vertices in the 
$150h^{-1}\hbox{Mpc}$ cubic volume involves a substantial 
degree of clustering, a clustering which is even more strongly borne out by 
the distribution of vertices in a thin slice through the box (bottom lefthand 
frame) and equally well reflected in the sky distribution (bottom righthand frame). 
The impression of strong clustering, on scales smaller than or of the order of 
the cellsize $\lambda_{\rm C}$, is most evidently expressed by the corresponding 
two-point correlation function $\xi(r)$ (Fig. 4, log-log plot lefthand frame, 
lin-lin plot in the righthand frame). Not only can we discern a
clear positive signal but -- surprising at the time of its finding on
the basis of similar computer experiments (van de Weygaert \& Icke 1989) --
out to a distance of at least $r \approx 1/4\,\lambda_{\rm c}$ the correlation
function appears to be an almost perfect power-law, 
\begin{equation}
\xi_{vv}(r)=\left({\displaystyle r_{\rm o} \over \displaystyle r}\right)
^\gamma\,;\hskip 2.0truecm \gamma = 1.95;\hskip 0.5truecm 
r_{\rm o} \approx 0.3\,\lambda_{\rm c}\,.
\end{equation}
The solid line in the log-log diagram in Fig. 4 represents the power-law with these 
parameters, the slope $\gamma \approx 1.95$ and ``clustering length'' $r_{\rm o} 
\approx 0.3\,\lambda_{\rm c}$. (the solid line represents the power-law with 
these parameters). Beyond 
this range, the power-law behaviour breaks down and following a
gradual decline the correlation function rapidly falls off to a zero value once
distances are of the order of (half) the cellsize. Assessing the behaviour of $\xi(r)$ 
in a linear-linear plot, we get a better idea of its behaviour around the 
zeropoint ``correlation length'' $r_{\rm a}\approx 0.5\lambda_c$ (bottom righthand frame 
fig. 4). Beyond $r_{\rm a}$ the distribution of Voronoi vertices
is practically uniform. Its only noteworthy behaviour is the gradually declining and 
alternating quasi-periodic ringing between positive and negative values similar to that 
we also recognized in the ``galaxy'' distribution, a vague echo of the
cellular patterns which the vertices trace out. Finally, beyond $r \approx 2 \lambda_c$ 
any noticeable correlation seems to be absent. 

The above 2pt correlation function of Voronoi vertices is a surprisingly good 
and solid match to the observed world. It sheds an alternative view on the 
power-law clustering with power law $\gamma \approx 2$ found in the cluster 
distribution. Also, the observed cluster clustering length $r_{\rm o} \approx 
20h^{-1}\,\hbox{Mpc}$ can be explained within the context of a cellular 
model, suggesting a cellsize of $\lambda_{\rm c} \approx 70h^{-1}\,\hbox{Mpc}$ 
as the basic scale of the cosmic foam.
\begin{figure}[t]
\vskip -0.75truecm
\centering\mbox{\hskip -0.2truecm\psfig{figure=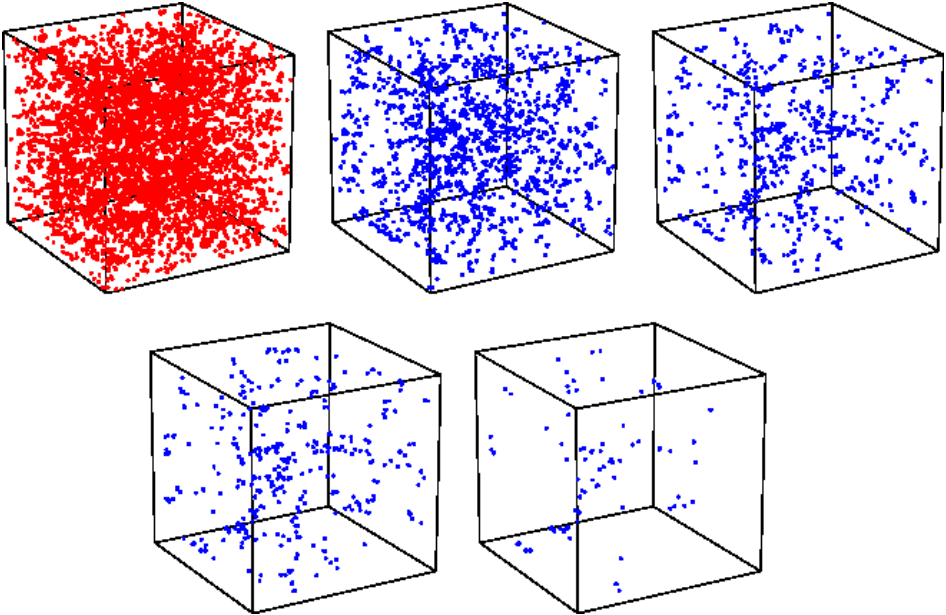,angle=270.,width=12.5cm}}
\caption{Selections of vertices from a full sample of vertices. Depicted are the 
(100$\%$) full sample (top left), and subsamples of the 25$\%$, 10$\%$, 5$\%$ and 
$1\%$ most massive vertices (top centre, top right, bottom left, bottom right). 
Note how the richer vertices appear to highlight ever more pronounced a 
filamentary superstructure running from the left box wall to the box centre. 
From: Van de Weygaert 2002a.}
\vskip -0.5truecm
\end{figure}
On the other hand, the latter also reveals a complication. The suggested cell scale 
is surely well in excess of the $25h^{-1}-35h^{-1}\,\hbox{Mpc}$ size of the
voids in the galaxy distribution. In addition, it appears to point to an 
internal inconsistency within the Voronoi concept. We saw above that 
if we tie the observed galaxy-galaxy correlation to the clustering of objects 
in the walls and filaments of the same tessellation framework, it suggests a 
cellsize $\lambda_c\approx 25h^{-1}\hbox{Mpc}$. This would conflict with 
the cellsize that would correspond to a good fit of the Voronoi vertex 
clustering to cluster clustering. The solution to this dilemma lead to an 
intriguing finding (for a complete description of this result see 
Van de Weygaert 2001). 

\subsection{Biased Voronoi Vertex Selections}
We first observe that the vertex correlation function of eqn.~(2) concerns
the full sample of vertices, irrespective of any possible selection
effects based on one or more relevant physical aspects. In reality, it will
be almost inevitable to invoke some sort of biasing through the definition 
criteria of the involved catalogue of clusters. Interpreting the Voronoi model in
its quality of asymptotic approximation to the galaxy distribution, its
vertices will automatically comprise a range of ``masses''. 

Brushing crudely over the details of the temporal evolution, we may assign each 
Voronoi vertex a ``mass'' estimate by equating that to the total amount of matter 
ultimately will flow towards that vertex. When we use the ``Voronoi streaming model'' as 
a reasonable description of the clustering process, it is reasonably straightforward 
if cumbersome to calculate the ``mass'' or ``richness'' ${\cal M}_{\rm V}$ of each
Voronoi vertex by pure geometric means. Evidently, vertices surrounding large cells 
are expected to be more massive. The details, turn out to be challengingly 
complex, as it concerns the (purely geometric) calculation of the volume of a 
non-convex polyhedron centered on the Voronoi vertex. The related nuclei
are the ones that supply the Voronoi vertex with inflowing matter.

To get an impression of the resulting selected vertex sets, Figure 5 shows 5 times 
the same box of $250h^{-1}\hbox{Mpc}$ size, each with a specific subset of the 
full vertex distribution (top lefthand cube). In the box we set up a realization of 
a Voronoi foam comprising 1000 cells with an average size of $25h^{-1}\hbox{Mpc}$. 
From the full vertex distribution we selected the ones whose ``richness'' 
${\cal M}_{\rm V}$ exceeds some specified lower limit. The depicted vertex subsets 
correspond to progressively higher lower mass limits, such that 100$\%$, 25$\%$, 
10$\%$, 5$\%$ and $1\%$ most massive vertices are included (from top lefthand to 
bottom righthand). The impression is not the one we would get if the subsamples 
would be mere random diluted subsamples from the full vertex sample. On the 
contrary, we get the definite impression of a growing coherence scale !!!  
Correcting for a possibly deceiving influence of the dilute sampling, and 
sampling an equal number of vertices from each ``selected'' sample only 
considerably strengthens this impression. There is an intrinsic effect 
in changing clustering properties as a function of (mass-defined) cluster sample. 
\begin{figure}[t]
\vskip -0.5truecm
\centering\mbox{\hskip -1.truecm\psfig{figure=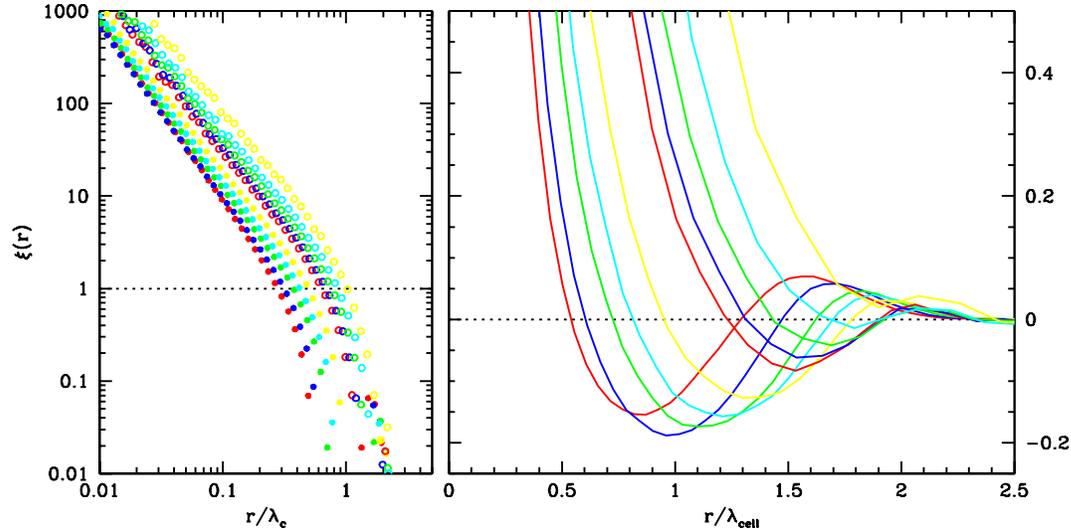,width=14.5cm}}
\vskip -0.5truecm
\caption{Scaling of the two-point correlation function of Voronoi vertices, for a 
variety of subsamples selected on the basis of ``richness'', ranging from samples 
with the complete population of vertices down to subsamples containing the 
$2.5\%$ most massive vertices. Left: log-log plot of $\xi(r)$ against 
$r/\lambda_c$, with $\lambda_c$ the basic tessellation cellsize 
($\equiv$ intranucleus distance). Notice the upward shift of $\xi(r)$ for  
subsamples with more massive vertices. Right: lin-lin plot of $\xi(r)$ against 
$r/\lambda_c$. Notice the striking rightward shift of the ``beating'' pattern as 
richness of the sample increases. From: Van de Weygaert 2002a.}
\vskip -0.3truecm
\end{figure}
\subsection{Vertex clustering: Geometric Biasing ?}
All in all, Fig. 5 provides ample testimony of a profound largely hidden large-scale 
pattern in foamlike networks, a hithero entirely unsuspected large-scale coherence 
over a range exceeding many cellsizes. 

To quantify the impression given by the distribution of the biased vertex selections, 
we analyzed the two-point correlation function for each vertex sample. We computed 
$\xi(r)$ for samples ranging from the complete sample down to the ones merely 
containing the $2.5\%$ most massive ones. As the average distance $\lambda_v(R) = 
n(R)^{-1/3}$ between the sample vertices increases monotonously with rising subsample 
richness, in the following we will frequently use the parameter $\lambda_v$ for  
characterizing the richness of the sample, ranging from $\lambda_{\rm v} 
\approx 0.5 \lambda_c$ up to $\lambda_{\rm v} \approx 1.5 \lambda_c$ for 
vertex samples comprising all vertices up to samples with the $10\%$ most massive 
vertices (the basic cellsize $\lambda_c$ functions objective distance unit). 

The surprising finding is that all subsamples of Voronoi 
vertices do retain a two-point correlation function displaying the same qualitative 
behaviour as the $\xi_{vv}(r)$ for the full unbiased vertex sample (Fig 6). Out 
to a certain range it invariably behaves like a power-law (lefthand frame), while 
beyond that range the correlation functions all show the decaying oscillatory 
behaviour that 
already has been encountered in the case of the full sample.
While all vertex $\xi_vv(r)$ convincingly confirm the impression of clustered point 
distributions, merely by the fact that it is rather straightforward to disentangle 
the various superposed two-point correlation functions we can immediately infer 
significant systematic differences. 
\begin{figure}[t]
\vskip -1.0truecm
\centering\mbox{\hskip -0.9truecm\psfig{figure=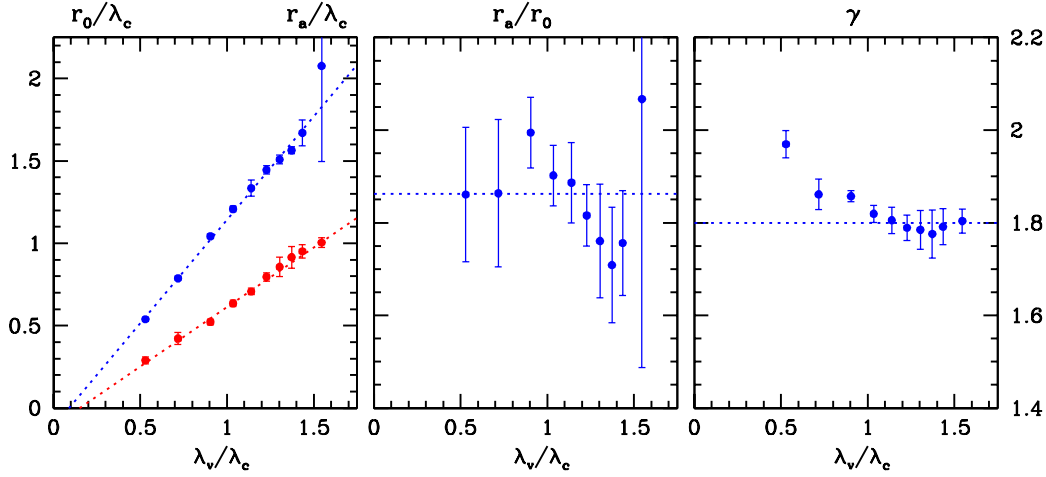,width=14.7cm}}
\vskip -7.4truecm
\caption{Scaling of Voronoi vertex two-point correlation function parameters for 
vertex subsamples over a range of ``richness''/``mass''. Left: 
the clustering length $r_0$ (red, $\xi(r_0) \equiv 1.0)$ and the correlation 
(coherence) length $r_a$ (blue, $\xi(r_a) \equiv 0$) as a function of average spatial 
separation between vertices in (mass) selected subsample, $\lambda_v/\lambda_c$. 
Centre: the ratio between clustering length $r_0$ and coherence length $r_a$ as 
function of subsample intravertex distance $\lambda_v/\lambda_c$. Right: the 
power-law slope $\gamma$ as function of $\lambda_v/\lambda_c$.}
\end{figure}
{\it First} observation is that the amplitude of the correlation functions 
increases monotonously with rising vertex sample richness. Expressing the 
amplitude in terms of the ``clustering length'' $r_{\rm o}$ and plotting this 
against the $\lambda_{\rm v}$ between 
the sample vertices (both in units of $\lambda_c$), a striking almost perfectly 
linear relation is resulting (Fig. 7, lefthand frame, lower line). In other words, almost 
out of the blue, the ``fractal'' clustering scaling description of Szalay \& Schramm 
(1985) appears to be stealthily hidden within foamy geometries. 
Although in the asymptotic Voronoi model we may be partially beset by 
the fact that we use an asymptotic measure for the vertex ``mass'' -- the total amount of 
mass that ultimately would settle in the nodes of the cosmic foam -- it may have 
disclosed that ultimately it reflects the foamy structured spatial matter distribution. 
Overall, the scaling of the clustering strength explains the impression of 
the increasingly compact clumpiness seen in the ``biased'' vertex distributions in 
Fig. 5. Summarizing, we can conclude that the foamy geometry is the ultimate ground for 
the observed amplified levels of cluster clustering. 

A {\it second} significant observation is that the lin-lin large-scale 
behaviour of the $\xi_{vv}$ seems to extend to larger and larger distances as the 
sample richness is increasing. The oscillatory behaviour is systematically shifting 
outward for the richer vertex samples, which reflects the fact that clustering patterns 
extend increasingly outward. Even though the basic cellular pattern 
had a characteristic scale of only $\lambda_c$, the sample of the $5\%$ richest 
nodes apparently seem to set up coherent patterns extending at least 2 to 3 times 
larger. This is clearly borne out by the earlier shown related point distributions 
(Fig. 5). Foamlike geometries seemingly induce coherent structures significantly larger 
than their basic size !!! This may hint at another tantalizing link between the 
galaxy and the cluster distribution. To elucidate this behaviour further, in Fig. 
7 (lefthand frame, higher line) we also plotted the ``correlation (coherence) scale'' 
$r_{\rm a}$ versus the average sample vertex distance $\lambda_v$. And yet again, 
as in the case of $r_{\rm o}$, we find an almost perfectly linear relation !!! 

Combining the behaviour of $r_{\rm o}$ and $r_{\rm a}$ we therefore find a 
remarkable `self-similar' scaling behaviour, in which the ratio of correlation 
versus clustering length is virtually constant for all vertex samples, 
$r_{\rm a}/r_{\rm o} \approx 1.86$ (see Fig. 7, central frame). {\it Foamlike 
networks appear to induce a clustering in which richer objects not only 
cluster more strongly, but also further out !!!} 

A {\it final} interesting detail on the vertex clustering scaling behaviour is that 
a slight and interesting trend in the behaviour of power-law slope. The richer 
samples correspond to a tilting of the slope. Interestingly, borne out by 
the lower righthand frame in Fig. 7, 
we see a gradual change from a slope $\gamma \approx 1.95$ for the full sample, 
to a robust $\gamma \approx 1.8$ for the selected samples. 
\begin{figure}[t]
\vskip -1.5truecm
\centering\mbox{\hskip -0.9truecm\psfig{figure=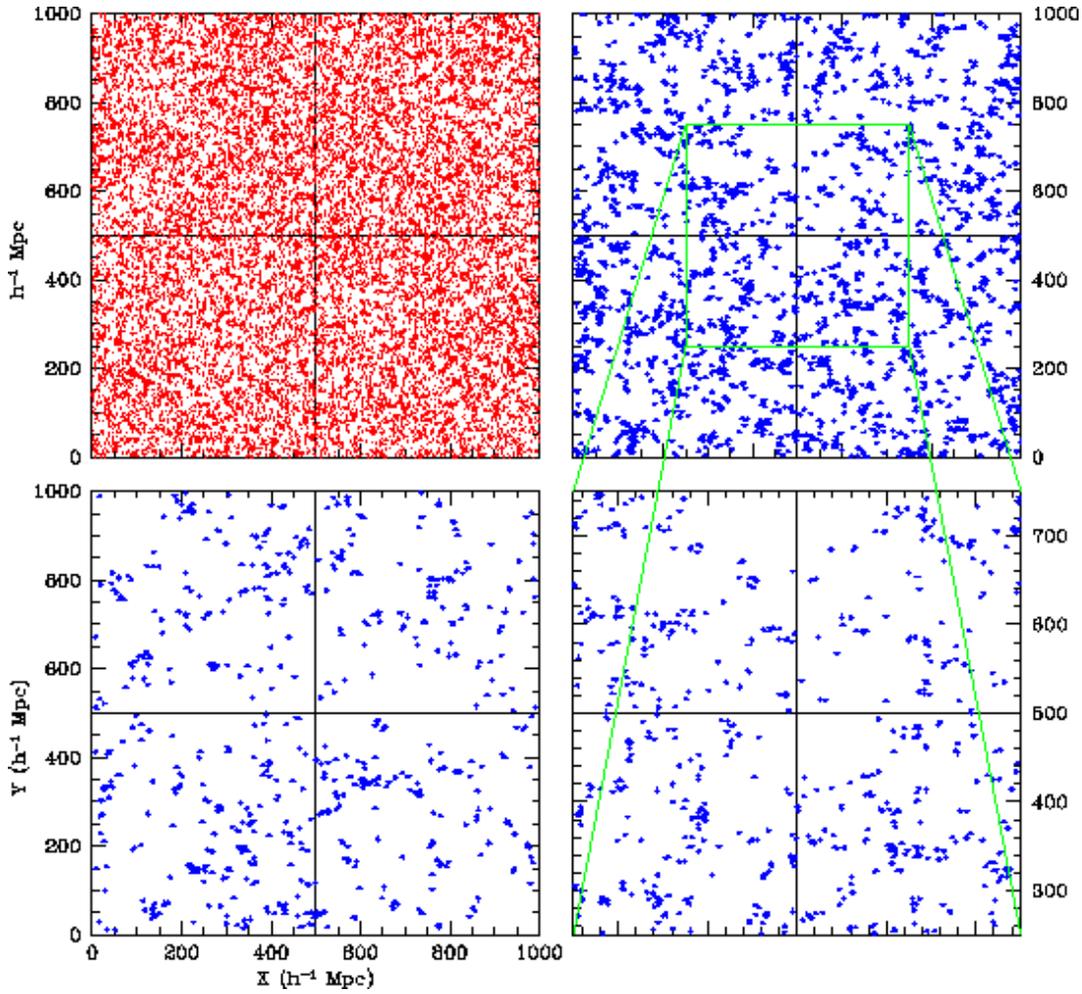,width=14.2cm}}
\vskip -0.3truecm
\caption{A depiction of the meaning of `self-similarity' in the vertex distribution. 
Out of a full sample of vertices (top left) in a central slice, (top right) the 
20.0$\%$ richest vertices. Similarly, (bottom left) the 2.5$\%$ richest vertices. 
When lifting the central 1/8$^{th}$ region out of the $20\%$ vertex subsample in 
the (top righthand) frame and sizing it up to the same scale as the full box, 
we observe the similarity in point process between the resulting (bottom 
righthand) distribution and that of the $2.5\%$ subsample (bottom lefthand). 
Self-similarity in pure form !}
\vskip -0.5truecm
\end{figure}
\section{Conclusions: Bias, Cosmic Geometry and Self-Similarity}
The uncovered systematic trends of vertex clustering have uncovered a hidden 
`self-similar' clustering of vertices. This may be appreciated best from 
studying a particular realization of such behaviour (see Fig. 8)

The above results form a tantalizing indication 
for the existence of self-similar clustering behaviour in spatial 
patterns with a cellular or foamlike morphology. It might hint at an 
intriguing and intimate relationship between the cosmic foamlike geometry 
and a variety of aspects of the spatial distribution of galaxies and 
clusters. One important implication is that with clusters residing at a
subset of nodes in the cosmic cellular framework, a configuration 
certainly reminiscent of the observed reality, it would explain 
why the level of clustering of clusters of galaxies becomes stronger 
as it concerns samples of more massive clusters. In addition, it would 
succesfully reproduce positive clustering of clusters over scales 
substantially exceeding the characteristic scale of voids and other 
elements of the cosmic foam. At these Megaparsec scales there is 
a close kinship between the measured galaxy-galaxy two-point 
correlation function and the foamlike morphology of the galaxy 
distribution. In other words, the cosmic geometry apparently 
implies a `geometrical biasing'' effect, qualitatively different 
from the more conventional ``peak biasing'' picture (Kaiser 1984). 
\section{Acknowledgments}
I would like to thank J. Dankers for his help in Voronoi 
plotting with Geomview. Writing 
this contribution, fond memories emerged of the many years over which the 
encouragement by Vincent Icke and Bernard Jones paved my way on the path 
through the world of Voronoi !

\end{document}